%% file: SystemicRiskInEurope.tex
\documentclass[a4paper,10pt]{article}
\usepackage[utf8]{inputenc}
\usepackage{graphicx}
\usepackage{booktabs}
\usepackage[numbers]{natbib}
\usepackage{amsmath}
\usepackage[top=2in, bottom=1.5in, left=3cm, right=3cm]{geometry}
\title{Cross-Correlations Between European Government Bonds And EuroStoxx
Assets}
\author{Jan Jurczyk, Alexander Eckrot}

\begin{document}
\maketitle
\begin{abstract}
We use principle component analysis (PCA) of cross correlations in European government bonds and
European stocks to investigate the systemic risk contained in the European economy.
We tackle the task to visualize the evolution of risk, introducing the conditional average rolling sum (CARS).
Using this tool we see that the risk of government bonds and stocks had an independent movement.
But in the course of the European sovereign debt crisis the coupling between bonds and stocks has strongly increased.
This results in an in-phase oscillation of risk for both markets since mid 2010.
In our data, we observe a steep amplitude increase, suggesting a high vulnerability of the two coupled systems.

\end{abstract}

\include{intro}
\include{methods}

\include{results}

\bibliographystyle{unsrtnat}
\bibliography{SystemicRiskInEurope}
\end{document}

%% file: intro.tex
\section{Introduction}
The inability of some European states to refinance their sovereign debts and 
the outbreak of the financial crisis in 2008 led in 2010 to the still ongoing European
sovereign debt crisis.
To stabilize the markets
and give incentives to invest in Europe, the European Union implemented 
the European Financial Stability Facility (EFSF) and later the European Stability 
Mechanism (ESM).
Heavily indebted European nations can use these tools to refinance their
government debt.\cite{esmreport2012} The utilization of this bailout support 
has to be accompanied by
fiscal consolidation, reforms and privatization of public goods.\cite{BLOOM}

Previous studies have shown, that principle component analysis (PCA) of cross-correlations in financial datasets 
is a wide field of study with many 
applications.\cite{laloux1999noise,gopikrishnan2001quantifying,plerou2002random,utsugi2004random,pan2007collective,
podobnik2010time,jiang2012anti,wang2013random,kwapien2006bulk,wilcox2004analysis}

More recent studies have shown that PCA is a viable tool to analyze the systemic risk, which measures the probability 
of events leading to widespread loss of confidence in the financial system 
\cite{zheng2012changes,billio2012econometric,kritzman2010principal,podobnik2010time, 
wang2011quantifying}. Zheng et al. have used the first eigenvalues
of the cross-correlation matrix between different stock-indices and stocks to 
reveal the rise in systemic risk, which lead to the financial crisis in 2007.

These studies show the importance of cross-correlation analysis for identifying the risk, but they lack a statement 
about the evolution of the systemic risk. For this purpose we introduce the conditional average rolling sum (CARS) of 
the eigenvalue time deviation, which allows us to track the systemic risk.
We apply this technique to a set of European government bonds and stocks listed in the EuroStoxx to examine the 
European sovereign debt crisis and show that since the beginning of the European dept crisis in 2010 these two 
markets are strongly correlated and both show the same oscillating risk progression.

%% file: methods.tex
\section{Method}
Detecting an increase of similar market reactions 
can be measured by utilizing the correlation matrix $C_{ij}$,
which contains the empirical correlation between two assets $i$ and $j$ in
a given time window $\tau$.

We study two different sets of time dependent assets:
The mean monthly returns $R_{i}=\partial_{t}\langle P(t)\rangle_{1m}$ of all stocks contained in the EuroStoxx50
and the change of the monthly yield return $R_{i}=\partial_{t}r_{i}(t)$ of 
the European 10 year bonds. We use logarithmic
returns 
\begin{equation}
\partial_{t}P(t)=ln\left(\frac{P(t)}{P(t-1)}\right)
\end{equation}
with the governmet bond data found on \cite{OECD} only allowed monthly values.

The correlation elements $C_{ij}$ are
\begin{equation}
 C_{ij}=\frac{\mathbf{E}\left[(R_{i}-\overline{R_{i}})(R_{j}-\overline{R_{j}})\right]}{\sigma_{i}\sigma_{j}}
\end{equation}
where $\sigma_{i}$ is the standard deviation of the returns of asset $i$.
Since the matrix is symmetric it is possible to use principle component analysis(PCA).
The eigenvalues ${\lambda_{1},...,\lambda_{N}}$ are called principal components of the orthogonal space, where the 
time-series are uncorrelated.
We track 
\begin{equation}
\tilde{\lambda}_{i}^{\Delta}(t)=\frac{\lambda_{i}^{\Delta}(t)}{N}
\end{equation}
where $\Delta$ is a chosen time window. In financial correlation matrices the first eigenvalue dominates, which 
shows that most of the information is contained in the largest eigenvalue and its eigenvector. Smaller eigenvalues are 
typically in the random regime and carry less information of the system, but are not pure 
noise.(\cite{wang2013random,kwapien2006bulk})

In the paper by Zheng et. al  \cite{zheng2012changes}, they observed that a steep increase in 
$\tilde{\lambda}_{i}^{\Delta}(t)$ is connected to 
the systemic risk in a set of assets. Peaks in the time derivation $\dot{\tilde{\lambda}}_{i}^{\Delta}(t)$ correspond 
to these events. 
An increasing risk can result in a crash. Thus we consider a build up process which improves the quality of 
identifying overall market moves. We take the conditional rolling sum of events
\begin{equation}
 \left\langle\dot{\lambda}\right\rangle_{\Omega}^{\dot{\tilde{\lambda}}(t)>0}=
\sum\limits_ 
{\omega\in\Omega}\frac{
1}{|\omega|}\sum\limits_{\substack{t\in\omega \\ 
\dot{\tilde{\lambda}}(t)>0}}\dot{\tilde{\lambda}}(t)=CARS
\end{equation}
which has a memory effect by default where
\begin{equation}
 \dot{\tilde{\lambda}}(t)=\frac{\partial}{\partial t}\left({\sum\limits_{i=1}^{4}\tilde{\lambda}_{i}^{\Delta}(t)}\right)
\end{equation}
takes the four largest eigenvalues into account in oder to project most of the information.

%% file: results.tex
\section{Results and Discussion}
\begin{figure}[!htbp]
  \begin{center}
    \begin{tabular}{cc}
    \resizebox{0.45\linewidth}{!}{\includegraphics{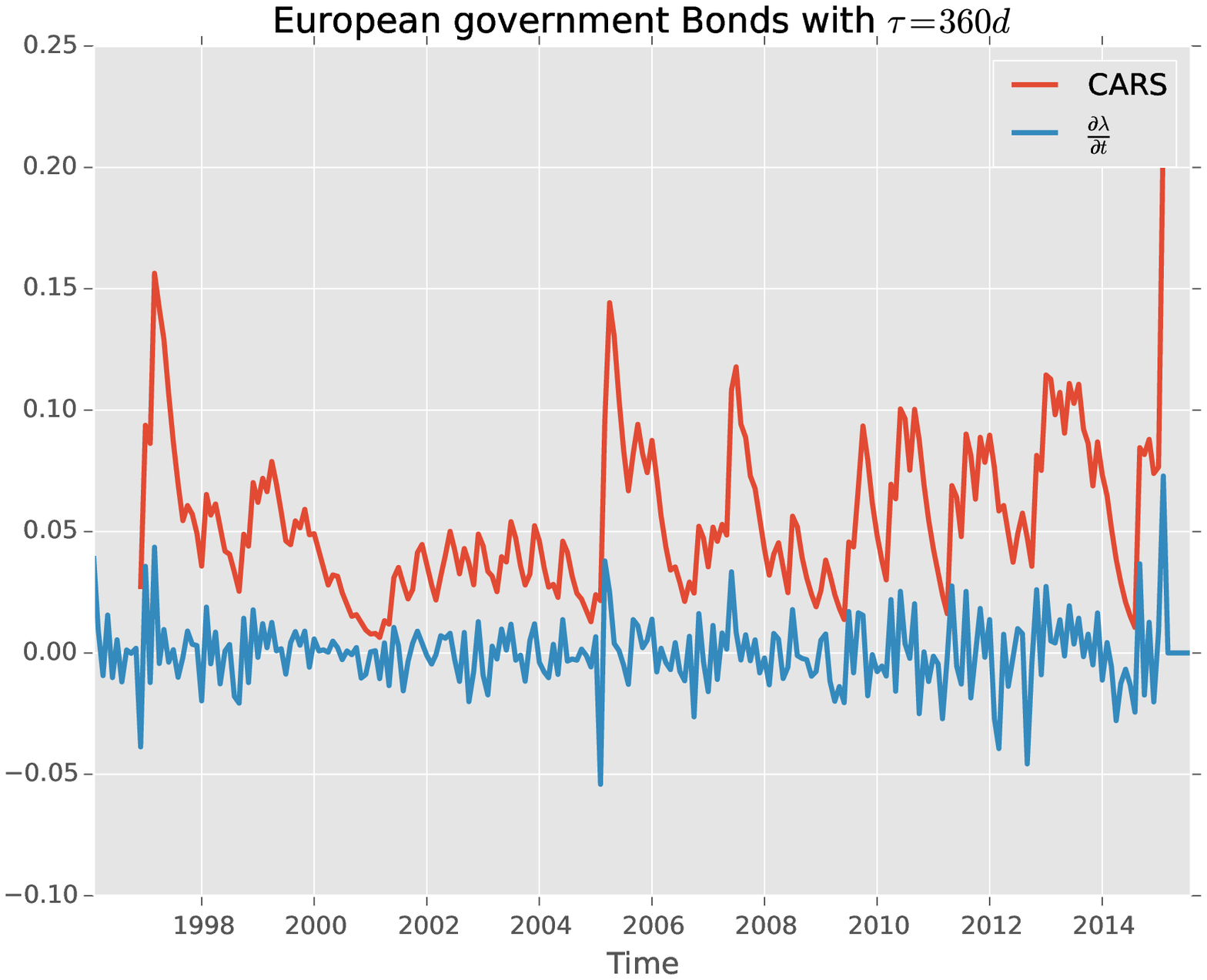}} &
      \resizebox{0.45\linewidth}{!}{\includegraphics{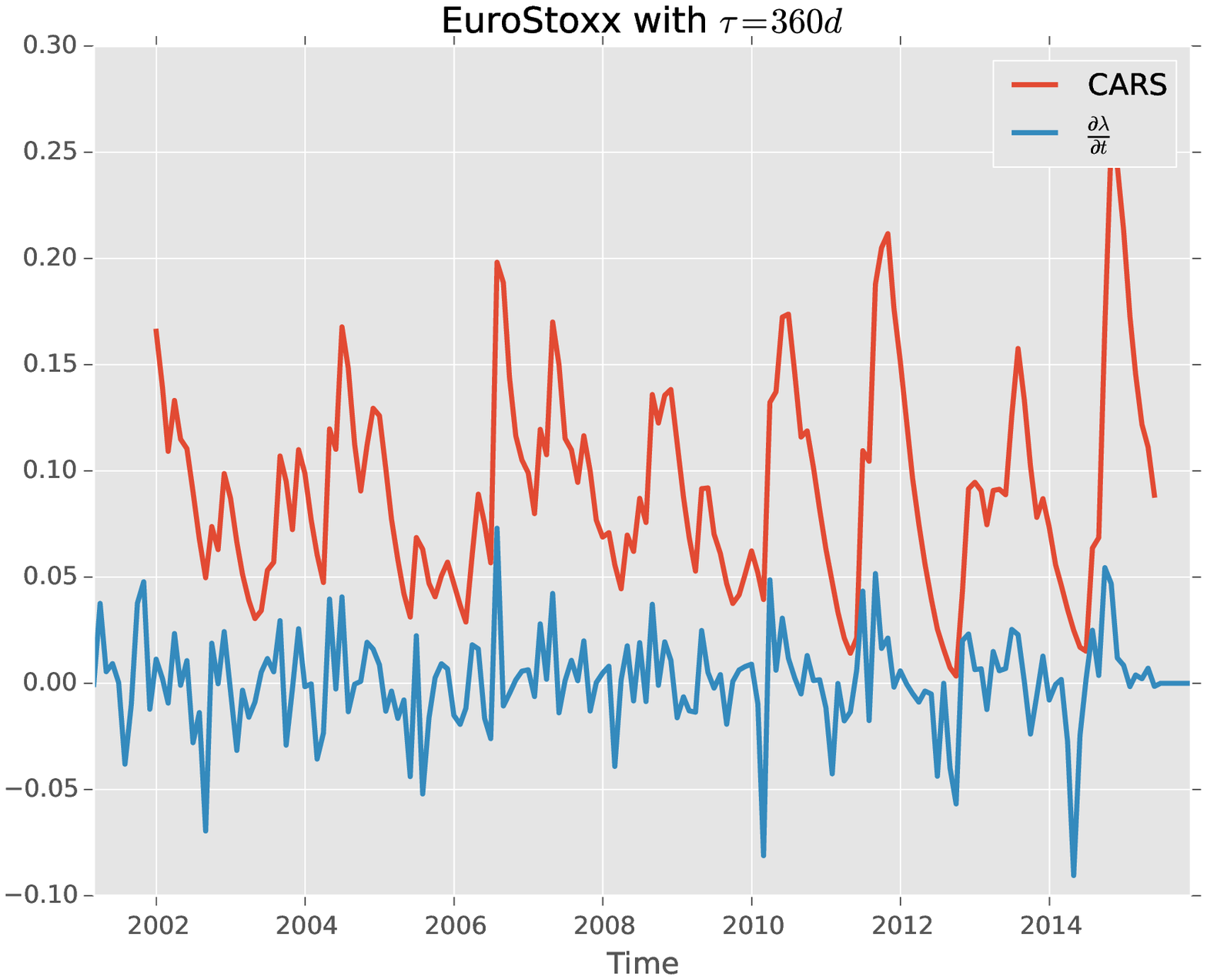}} \\
      \resizebox{0.45\linewidth}{!}{\includegraphics{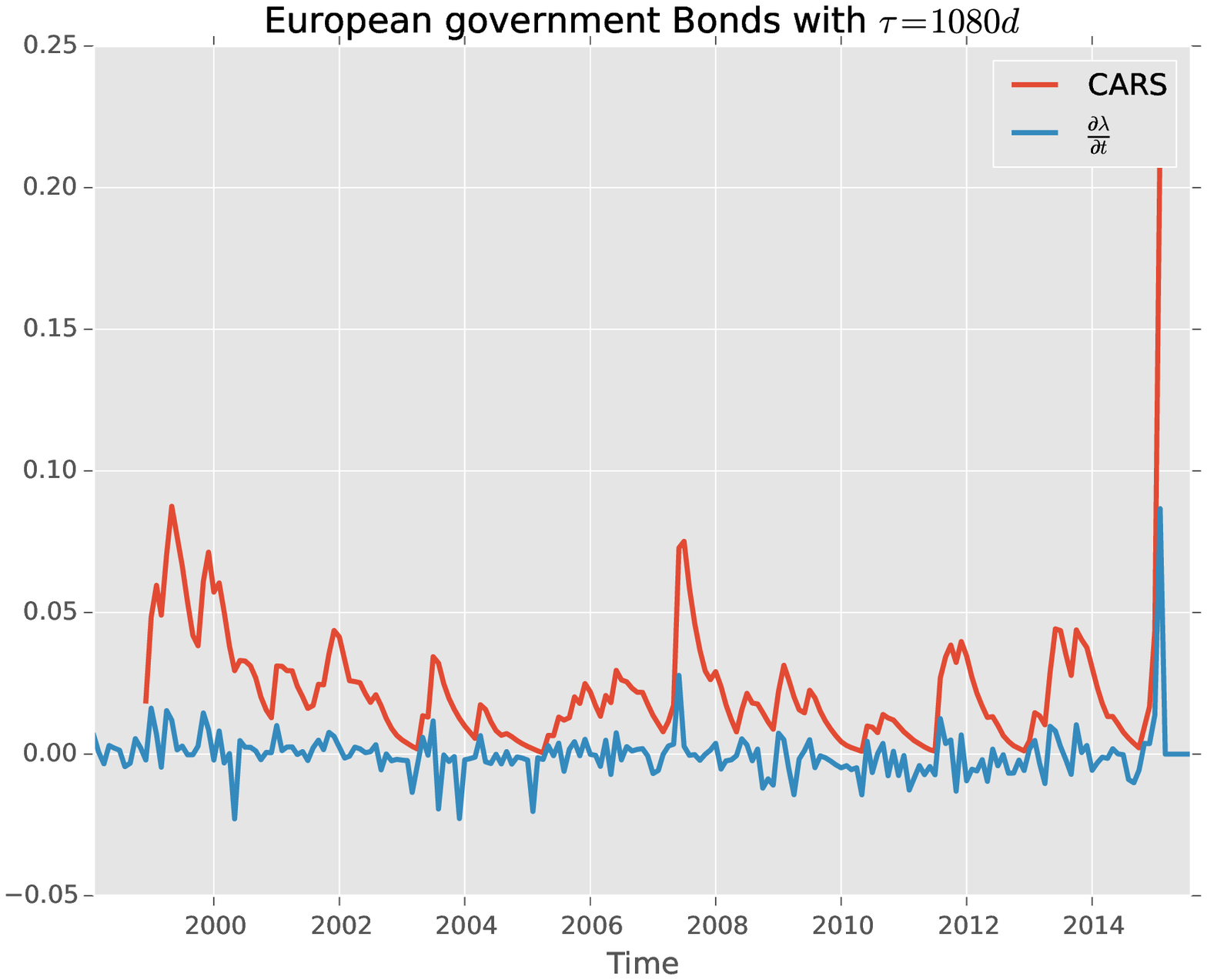}} &
      \resizebox{0.45\linewidth}{!}{\includegraphics{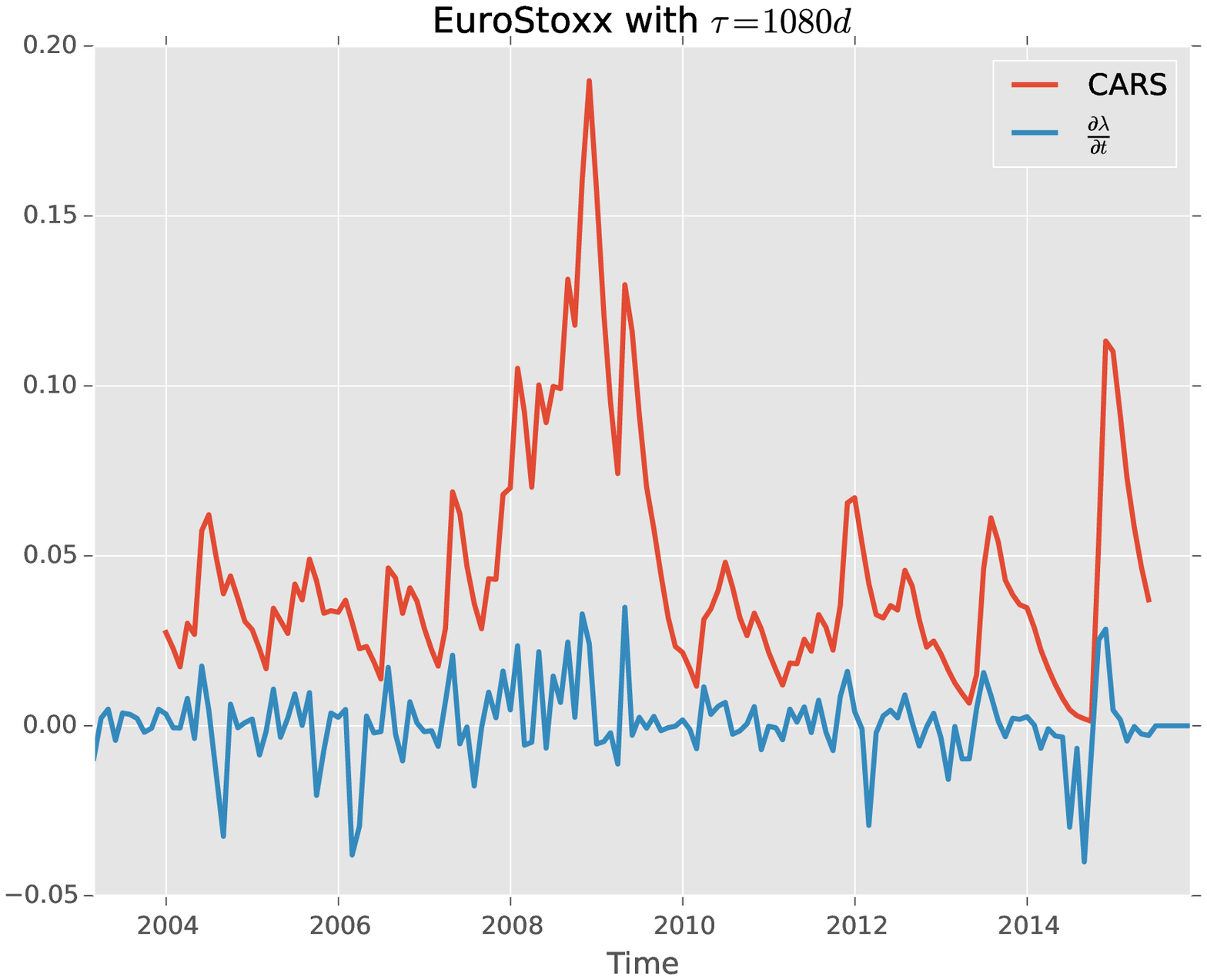}} \\
    \end{tabular}
  \end{center}
  \caption{Changes in the largest eigenvalues of cross-correlations and corresponding CARS.Top 
    panels use a time window of $\tau=1y$ while bottom panels use $\tau=3y$. The left side corresponds to
    European government bonds and the right side to the assets covered by the EuroStoxx. In the $\tau=1y$ panels 
    we observe CARS oscillations since June 2010. The $\tau=3y$ graphs show peaks with a high reliability in terms of 
    crisis forecasting.}
  \label{fig:timeVsCars}
\end{figure}

Applying PCA and the conditional average rolling sum (CARS) with a time window of 36 months on the EuroStoxx data 
(bottom right panel of figure \ref{fig:timeVsCars}),
we see a peak in November 2008. 
One can observe, that this peak is the consequence of many events in the derivation 
$\frac{\partial}{\partial t}\lambda$ between 
2007 and 2009,
and thus connected with the financial crisis. Furthermore we see two minor peaks in November
2011 and August 2013, which are associated with the European sovereign debt crisis.

\begin{figure}[!htbp]
  \begin{center}
    \begin{tabular}{cc}
    \resizebox{0.45\linewidth}{!}{\includegraphics{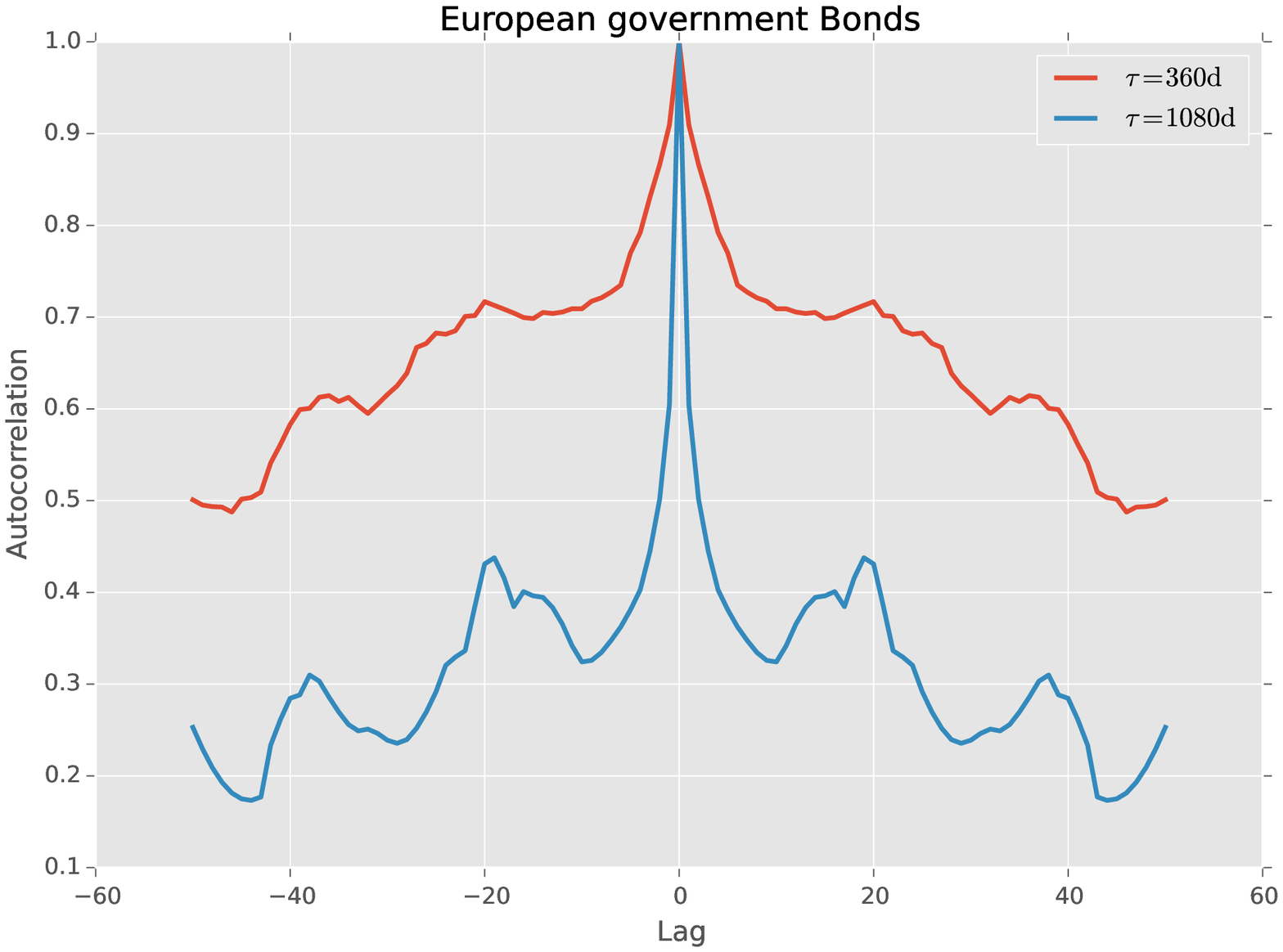}}&
    \resizebox{0.45\linewidth}{!}{\includegraphics{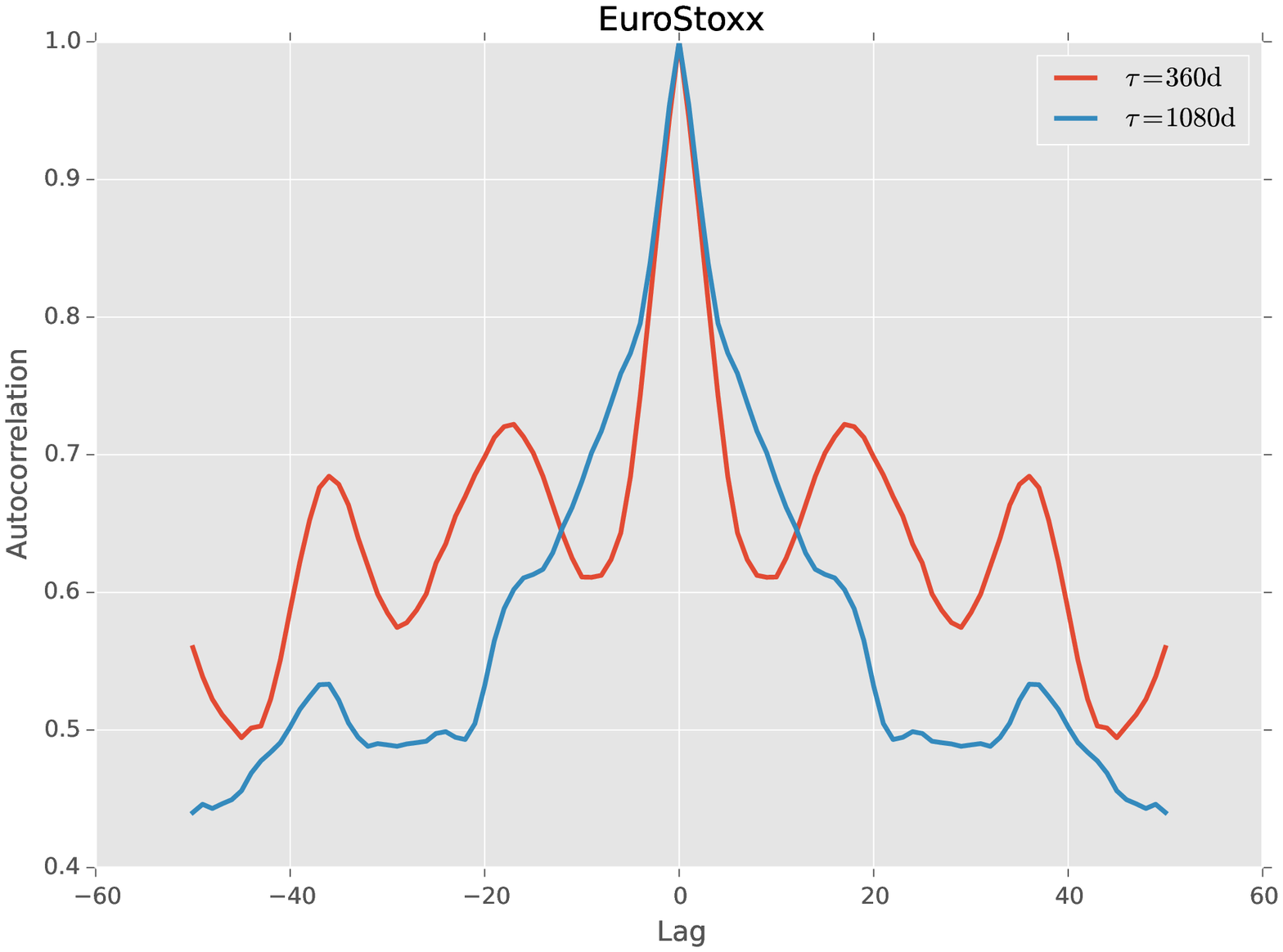}}\\
    \end{tabular}  
  \end{center}
  \caption{The autocorrelation of European government bonds (left) and EuroStoxx assets (right) with a time-window 
$\tau=1y,3y$. All cases exhibit an oscillation with frequency $\nu_{EuroBonds}^{1y,3y}\approx(20m)^{-1}$ and 
$\nu_{EuroStoxx}^{1y,3y}\approx(17m)^{-1}$}
  \label{fig:autocorrelation}
\end{figure}

Performing the same analysis for European long term government bonds, we see a very similar pattern in the bottom left 
panel of figure \ref{fig:timeVsCars}:
A high peak in July 2007, which is at the beginning of the build-up process for the broad peak in the EuroStoxx.
This peak is followed by smaller ones in December 2011 and August 2013. These dates correspond well with
the analysis of the stocks cross-correlations and show a strong connection in the risk of European stocks
and European government bonds.
At the end of 2015 we detect for both graphs a distinct increase in the first eigenvalues of the cross-correlation 
matrix. Especially for the government bonds this increase reaches unobserved high values.
The top panels of figure \ref{fig:timeVsCars} show the CARS of the European government bonds (left) and EuroStoxx 
assets (right)  for time-window $\tau=1y$. Choosing this time-window results in a fluctuating pattern, which is not as 
suitable for indicating big market changes as the $\tau=3y$ but it clearly features the capability of discovering 
oscillation patterns in the CARS.
We can find these oscillations in all four cases, indicating
that the risk in these markets oscillates on a time frame of several months. 
While these two oscillations have been out of phase before, they are in phase since 2010.
In figure \ref{fig:autocorrelation} we show the periodic property by plotting the the autocorrelation function for both 
datasets.

\begin{figure}[!htbp]
  \begin{center}
    \begin{tabular}{cc}
    \resizebox{0.45\linewidth}{!}{\includegraphics{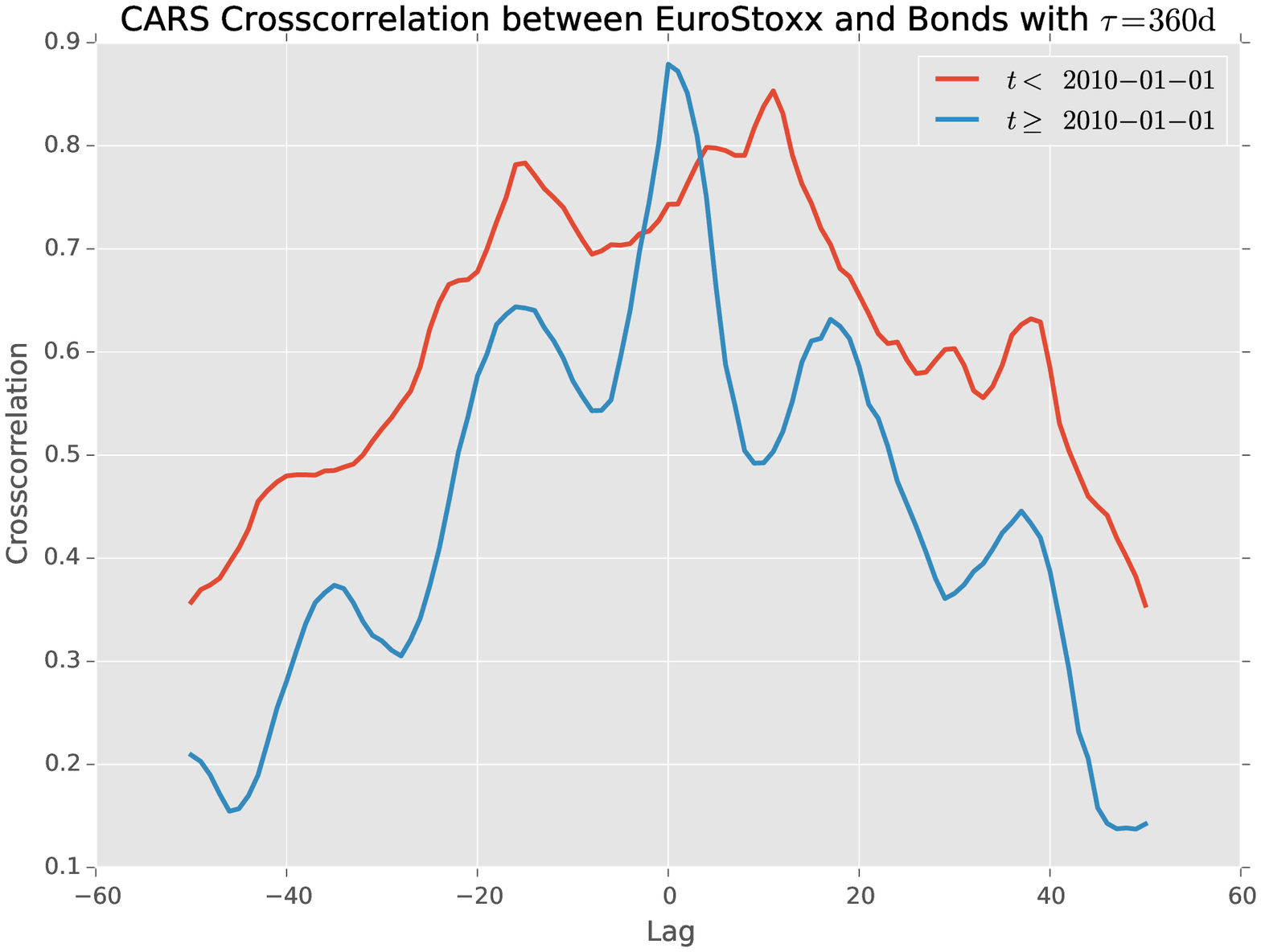}}&
    \resizebox{0.45\linewidth}{!}{\includegraphics{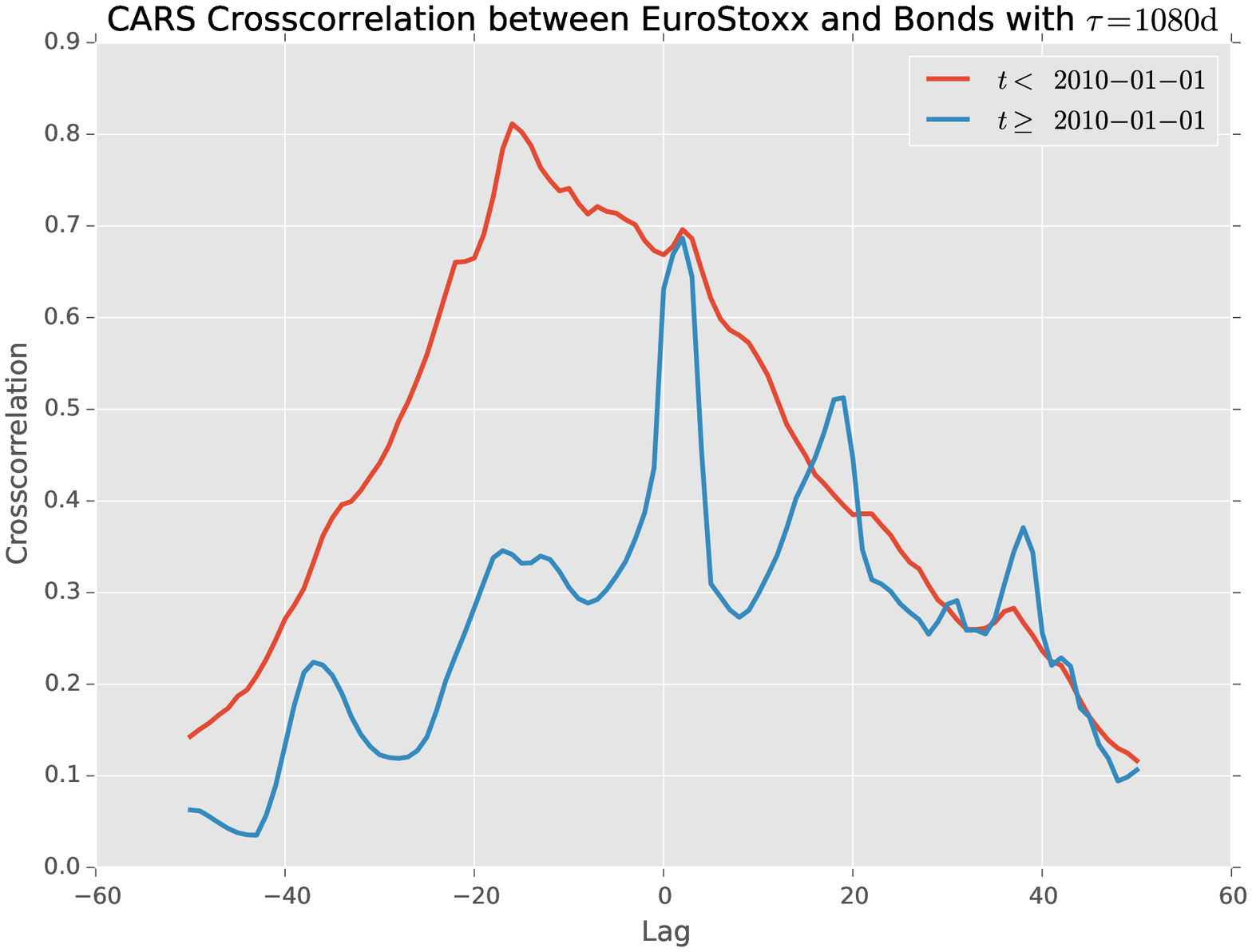}}  \\
    \end{tabular}
  	\end{center}
    \caption{Cars cross-correlation between  European government bonds and EuroStoxx assets before and after 2010. The 
      maximum cross-correlation in the $\tau=1y,3y$ panels both shift from a non zero lag before 2010 to a approx.
      zero month lag after 2010. Also both time-windows feature an oscillation with $\nu\approx (16m)^{-1},(19m)^{-1}$ 
      after 2010.}
  \label{fig:crosscorrelation}
\end{figure}

We emphasize this fact in figure \ref{fig:crosscorrelation}, which shows the correlations between the two CARS  
before and after 2010.

Before 2010 we find that the maximum 
correlation has a non zero lag, indicating only weak interactions between these two markets. Looking at the 
cross-correlation after 2010, we measure a zero time lag for the maximum correlation in the case of $\tau=1y$ and an 
oscillation . 
This fact suggests that these two market are strongly coupled since 2010.

\section{Conclusion}

In this article we have shown, that conditional average rolling sum of events is a suitable tool for the visualization 
of risks. The in-build memory ensures that past events are taken into account in order to track time-dependent 
processes. A clear connection between a high CARS value and an high systemic risk has been established.

Furthermore we observed an oscillation in the stock and government bonds markets. Originally these oscillations in 
those two markets were not coupled until 2010. From this point on we observe a zero or one month time-lag in the 
cross-correlations between them.

In November 2014 an unprecedented surge in the conditional average rolling sum occurred. It seems that the sharp risk 
increase manifested in the so called ``Schnitzel-Crisis'' in May 2015.

%% file: SystemicRiskInEurope.bbl
\begin{thebibliography}{17}
\providecommand{\natexlab}[1]{#1}
\providecommand{\url}[1]{\texttt{#1}}
\expandafter\ifx\csname urlstyle\endcsname\relax
  \providecommand{\doi}[1]{doi: #1}\else
  \providecommand{\doi}{doi: \begingroup \urlstyle{rm}\Url}\fi

\bibitem[ESM(2012)]{esmreport2012}
ESM.
\newblock {ESM Annual Report 2012}.
\newblock Online, 2012.

\bibitem[News(2011)]{BLOOM}
Bloomberg News.
\newblock {European Crisis Timeline From Maastricht to Papandreou Exit },
  November 2011.
\newblock Details on the timeline can be found here:
  http://www.bloomberg.com/news/articles/2011-11-07/europe-timeline-maastricht-to-papandreou.

\bibitem[Laloux et~al.(1999)Laloux, Cizeau, Bouchaud, and
  Potters]{laloux1999noise}
Laurent Laloux, Pierre Cizeau, Jean-Philippe Bouchaud, and Marc Potters.
\newblock {Noise dressing of financial correlation matrices}.
\newblock \emph{Physical review letters}, 83\penalty0 (7):\penalty0 1467, 1999.

\bibitem[Gopikrishnan et~al.(2001)Gopikrishnan, Rosenow, Plerou, and
  Stanley]{gopikrishnan2001quantifying}
Parameswaran Gopikrishnan, Bernd Rosenow, Vasiliki Plerou, and H~Eugene
  Stanley.
\newblock {Quantifying and interpreting collective behavior in financial
  markets}.
\newblock \emph{Physical Review E}, 64\penalty0 (3):\penalty0 035106, 2001.

\bibitem[Plerou et~al.(2002)Plerou, Gopikrishnan, Rosenow, Amaral, Guhr, and
  Stanley]{plerou2002random}
Vasiliki Plerou, Parameswaran Gopikrishnan, Bernd Rosenow, Luis A~Nunes Amaral,
  Thomas Guhr, and H~Eugene Stanley.
\newblock {Random matrix approach to cross correlations in financial data}.
\newblock \emph{Physical Review E}, 65\penalty0 (6):\penalty0 066126, 2002.

\bibitem[Utsugi et~al.(2004)Utsugi, Ino, and Oshikawa]{utsugi2004random}
Akihiko Utsugi, Kazusumi Ino, and Masaki Oshikawa.
\newblock {Random matrix theory analysis of cross correlations in financial
  markets}.
\newblock \emph{Physical Review E}, 70\penalty0 (2):\penalty0 026110, 2004.

\bibitem[Pan and Sinha(2007)]{pan2007collective}
Raj~Kumar Pan and Sitabhra Sinha.
\newblock {Collective behavior of stock price movements in an emerging market}.
\newblock \emph{Physical Review E}, 76\penalty0 (4):\penalty0 046116, 2007.

\bibitem[Podobnik et~al.(2010)Podobnik, Wang, Horvatic, Grosse, and
  Stanley]{podobnik2010time}
Boris Podobnik, Duan Wang, Davor Horvatic, Ivo Grosse, and H~Eugene Stanley.
\newblock {Time-lag cross-correlations in collective phenomena}.
\newblock \emph{EPL (Europhysics Letters)}, 90\penalty0 (6):\penalty0 68001,
  2010.

\bibitem[Jiang and Zheng(2012)]{jiang2012anti}
XF~Jiang and B~Zheng.
\newblock {Anti-correlation and subsector structure in financial systems}.
\newblock \emph{EPL (Europhysics Letters)}, 97\penalty0 (4):\penalty0 48006,
  2012.

\bibitem[Wang et~al.(2013)Wang, Xie, Chen, Yang, and Yang]{wang2013random}
Gang-Jin Wang, Chi Xie, Shou Chen, Jiao-Jiao Yang, and Ming-Yan Yang.
\newblock {Random matrix theory analysis of cross-correlations in the US stock
  market: Evidence from Pearson{\rq}s correlation coefficient and detrended
  cross-correlation coefficient}.
\newblock \emph{Physica A: Statistical Mechanics and its Applications},
  392\penalty0 (17):\penalty0 3715--3730, 2013.

\bibitem[Kwapie{\'n} et~al.(2006)Kwapie{\'n}, Dro{\.z}d{\.z}, O{\'s}wie,
  et~al.]{kwapien2006bulk}
J~Kwapie{\'n}, S~Dro{\.z}d{\.z}, P~O{\'s}wie, et~al.
\newblock {The bulk of the stock market correlation matrix is not pure noise}.
\newblock \emph{Physica A: Statistical Mechanics and its Applications},
  359:\penalty0 589--606, 2006.

\bibitem[Wilcox and Gebbie(2004)]{wilcox2004analysis}
Diane Wilcox and Tim Gebbie.
\newblock {On the analysis of cross-correlations in South African market data}.
\newblock \emph{Physica A: Statistical Mechanics and its Applications},
  344\penalty0 (1):\penalty0 294--298, 2004.

\bibitem[Zheng et~al.(2012)Zheng, Podobnik, Feng, and Li]{zheng2012changes}
Zeyu Zheng, Boris Podobnik, Ling Feng, and Baowen Li.
\newblock {Changes in cross-correlations as an indicator for systemic risk}.
\newblock \emph{Scientific reports}, 2, 2012.

\bibitem[Billio et~al.(2012)Billio, Getmansky, Lo, and
  Pelizzon]{billio2012econometric}
Monica Billio, Mila Getmansky, Andrew~W Lo, and Loriana Pelizzon.
\newblock {Econometric measures of connectedness and systemic risk in the
  finance and insurance sectors}.
\newblock \emph{Journal of Financial Economics}, 104\penalty0 (3):\penalty0
  535--559, 2012.

\bibitem[Kritzman et~al.(2010)Kritzman, Li, Page, and
  Rigobon]{kritzman2010principal}
Mark Kritzman, Yuanzhen Li, Sebastien Page, and Roberto Rigobon.
\newblock {Principal components as a measure of systemic risk}.
\newblock 2010.

\bibitem[Wang et~al.(2011)Wang, Podobnik, Horvati{\'c}, and
  Stanley]{wang2011quantifying}
Duan Wang, Boris Podobnik, Davor Horvati{\'c}, and H~Eugene Stanley.
\newblock {Quantifying and modeling long-range cross correlations in multiple
  time series with applications to world stock indices}.
\newblock \emph{Physical Review E}, 83\penalty0 (4):\penalty0 046121, 2011.

\bibitem[OECD(2015)]{OECD}
OECD.
\newblock {Main Economic Indicators - complete database}.
\newblock February 2015.
\newblock \doi{10.1787/data-00052-en}.

\end{thebibliography}
